# Exploring WorldCat Identities as an altmetric information source: A library catalog analysis experiment in the field of Scientometrics

Daniel Torres-Salinas[1], Wenceslao Arroyo-Machado[2] & Mike Thelwall[3]

[1]Department of Information and Communication Sciences, University of Granada, Faculty of Communication and Documentation, Granada, Spain; 0000-0001-8790-3314

[2] Department of Information and Communication Sciences, University of Granada, Faculty of Communication and Documentation, Granada, Spain; 0000-0001-9437-8757; wences@ugr.es [corresponding author]

[3]Statistical Cybermetrics Research Group (SCRG), University of Wolverhampton, Wulfruna Street, Wolverhampton, WV1 1LY, UK; 0000-0001-6065-205X

**Abstract**

Assessing the impact of scholarly books is a difficult research evaluation problem. Library Catalog Analysis facilitates the quantitative study, at different levels, of the impact and diffusion of academic books based on data about their availability in libraries. The WorldCat global catalog collates data on library holdings, offering a range of tools including the novel WorldCat Identities. This is based on author profiles and provides indicators relating to the availability of their books in library catalogs. Here, we investigate this new tool to identify its strengths and weaknesses based on a sample of Bibliometrics and Scientometrics authors. We review the problems that this entails and compare Library Catalog Analysis indicators with Google Scholar and Web of Science citations. The results show that WorldCat Identities can be a useful tool for book impact assessment but the value of its data is undermined by the provision of massive collections of ebooks to academic libraries.

**Declarations**

**Funding (information that explains whether and by whom the research was supported)**
This work has been possible thanks to financial support from "InfluScience - Scientists with social influence: a model to measure knowledge transfer in the digital society" (PID2019-109127RB-I00 / SRA /10.13039/501100011033), a project funded by scientific research team grants from the Ministry of Science and Innovation of Spain.

**Availability of data and material (data transparency)**
All data are available at https://github.com/Wences91/library_catalog_wi/

**Code availability**
All the code with the data processing is available at https://github.com/Wences91/library_catalog_wi/

**Alternative DOI:** https://doi.org/10.5281/zenodo.4288623 (ZENODO)

**Authors' contributions**
All authors contributed to the study conception and design. Material preparation and data collection was carried out by Wenceslao Arroyo-Machado. Analysis were performed by Daniel Torres-Salinas and Mike Thelwall. The first draft of the manuscript was written by Daniel Torres-Salinas and all authors commented on previous versions of the manuscript. All authors read and approved the final manuscript.

**Acknowledgments**
This work has been possible thanks to financial support from "InfluScience - Scientists with social influence: a model to measure knowledge transfer in the digital society" (PID2019-109127RB-I00 / SRA /10.13039/501100011033), a project funded by scientific research team grants from the Ministry of Science and Innovation of Spain. Daniel Torres-Salinas has received funding from the University of Granada's "Plan Propio de Investigación y Transferencia" under the "Reincorporación de Jóvenes Doctores" grant. Wenceslao Arroyo-Machado has received funding from the Spanish Ministry of Universities under the FPU Grant (FPU18/05835).

# 1. Introduction

The importance of books and monographs in scientific communication has been recognized for a long time (Archambault, Vignola-Gagné, Côté, Larivière & Gingrasb, 2006; Hicks, 1999; Huang & Chang, 2008). Early bibliometric impact evaluations of books were restricted to the limited data available in citation indexes. A key problem was that citation analysis databases primarily indexed journal articles, with limited coverage of books. This was addressed by new indicators based on library catalogs that became universally accessible through the Z39.50 protocol for internet-based search/retrieval and the launch of WorldCat.org, which used Z39.50 to collate library holdings from all over the world. The WorldCat.org open-access catalog unified millions of libraries in a single search engine enabling users to count how many libraries contained any given book, creating an alternative type of impact evidence (Nilges, 2006).

The library count method was termed Library Catalog Analysis (LCA) (Torres-Salinas & Moed, 2008) or Library Holdings Analysis (Linmans, 2008) and challenged the use of traditional citations two years before the publication of the Altmetric manifesto suggested that social media mentions could be used to track the societal impacts of academic publications (Priem, Taraborelli, Groth & Neylon, 2010). Since then, many other methods of analyzing book diffusion have appeared. These include the number of reviews recorded by the Book Review Index or the number and score available on the Goodreads or Amazon Reviews websites—the latter being related to popularity (Kousha & Thelwall, 2016). Similarly, in recent years, mentions in course syllabi, the Mendeley social reference sharing site, or YouTube comments (Kousha & Thelwall, 2015) have also been used.

In 2009, LCA was defined as "the application of bibliometric techniques to a set of library online catalogs" and in a case study, Torres-Salinas and Moed (2008, 2009) selected the field of Economics and analyzed 121 147 titles included on 417 033 occasions in 42 libraries. Similarly, White et al. (2009) proposed an identical method for what they termed "libcitations" that focused on the micro-level: 148 authors from different departments (Philosophy, History and Political Science) at two Australian universities (New South Wales, Sydney). They used WorldCat as their source of information. These studies showed that LCA was practical and gave plausible results.

Libcitations offer an alternative vision of the impact of books, with correlations with citations usually being low (Linmans, 2010; Zuccala & Guns, 2013; Zhang, Zhou & Zhang, 2018). Linmans' data set generated a correlation of 0.29, rising to 0.49 for English language books. Zuccala obtained correlations of 0.24 and 0.20 for History and Literature, respectively. Kousha and Thelwall (2016) compared libcitation correlations with various Amazon indicators. Their strongest correlation with Amazon Reviews was 0.348 for the humanities. Torres-Salinas, Gumpenberger and Gorraiz (2017) also reported a low correlation between library counts and other altmetrics included in PlumX Analytics.

WorldCat dominates library catalog studies (e.g., Linmans, 2010; Zuccala & White, 2015; Neville & Henry, 2014; Halevi, Nicolas, & Bar-Ilan, 2016) partly because of its uniquely large size. The OCLC directory currently identifies 15 194 libraries, 5804 of which are academic[1]. PlumX, currently owned by Elsevier, includes WorldCat library holdings (Holdings: WorldCat) among its altmetric indicators and facilitates both their calculation through lSBNs and large-scale searches (Torres-Salinas, Robinson-Garcia, Gorraiz, 2017). One of the first studies to use this source was Halevi, Nicolas and Bar-Ilan (2016), which used 71 443 eBook ISBN numbers.

WorldCat offers a wide range of functions in addition to its core index, such as the experimental WorldCat Identities (WI). This tool brings together the "complete works" of any given author, reporting the library diffusion data of their works overall for the author and by individual publication. It also integrates context-based data (e.g., genre, topics, name variants, co-authors) (Fig. 1). WI therefore provides Library Catalog Metrics profiles similar to that of other academic sites (Google Scholar Profiles, ResearchGate or custom current research information systems (CRIS) at research institutions), except WI is based mainly on books and bibliographic records indexed in library catalogs. These profiles open an interesting methodological door because library holdings have previously been studied at the level of record or work rather than aggregated by author.

---

[1] Information drawn from the Directory of OCLC Members: https://www.oclc.org/en/contacts/libraries.html. Note that some OCLC sources put the number of member libraries at 17 983: https://www.oclc.org/en/about.html

**Fig. 1** Basic information offered by WorldCat Identities for a given author

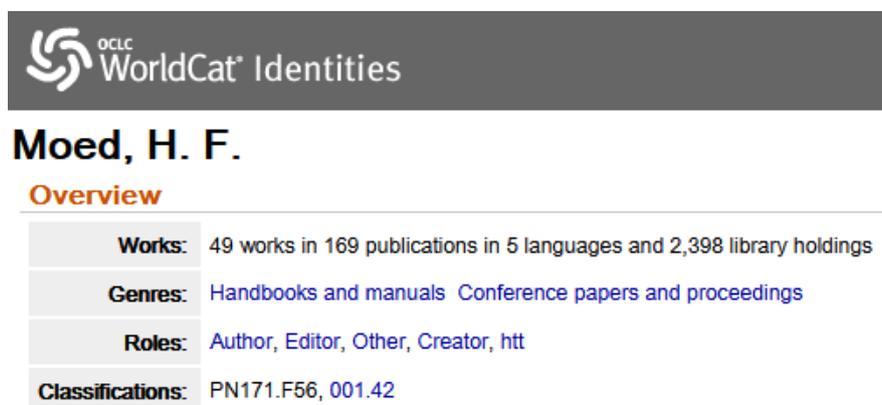

Despite the potential usefulness of WI, no study of its use has been undertaken from a metric perspective. Consequently, there is a need to test the value of WI as a source of information to obtain indicators based on library catalog for authors. Furthermore, WI includes information drawn from other sources and uses unspecified methods—presumably computational—to match works to authors[2]. Although other studies have used library holdings at the author level to a limited extent (White & Zuccala, 2018), no prior study has analyzed the authors in a specific scientific specialty. Here we address this gap for the field of Scientometrics and its most frequently cited authors as a sample, with the following objectives:

1) To analyze the strengths and weaknesses of WI as a source of information about the presence of an author's works in library catalogs.
2) To assess the value of LCA for describing a scientific field by analyzing WI indicators for the library holdings of scientometricians.
    2.1) To compare citation indicators and indicators based on library holdings at the author level.
    2.2) To identify the most widely distributed library works in a given scientific field from WI author profiles.

This paper is an extension of a book chapter analysing Informetrics authors (Torres-Salinas & Arroyo-Machado, 2020). It is organized as follows: first we describe the methodological process of identifying authors and gathering WI data. Secondly, we present results on the number of library holdings of Informetrics researchers (author level analysis) and compare these with Google Scholar citations. We also apply LCA to determine which books it identifies as being the most important (book level analysis). Finally, we discuss our results, acknowledging their limitations, and stressing the potential use of WI in the analysis of any specific field.

## 2. Methodology

### 2.1. Selecting our sample of researchers

To apply WI to a set of researchers who specialize in Scientometrics, we selected a sample of authors included in the "Scholar Mirrors" portal[3]—which gathers profiles of researchers in Bibliometrics, Scientometrics, Informetrics, Webometrics, and Altmetrics on platforms like Google Scholar or ResearchGate. The "Scholar

---

[2] https://www.oclc.org/research/areas/data-science/identities.html
[3] Scholar Mirrors: [2020-02-11]: http://www.scholar-mirrors.infoec3.es/

Mirrors" was created in 2015 by the EC3 Group at the University of Granada (Spain) and has identified a total of 813 Google profiles of which 398 are classified as core authors[4]. We selected these 398 authors for our study to represent a large sample of active authors in the broad field of Scientometrics.

## 2.2. WorldCat Identities

The WI author profile provides information about an author's bibliographic production. It includes the following four indicators (Fig. 1):

- Works: Number of different works indexed in WorldCat.
- Publications: Total number of works indexed in WorldCat, including separate book editions.
- Languages: Number of different languages in which an author's works, including different editions, have been published.
- Library holdings: Number of different WorldCat member libraries that hold the author's works.

Table 1 shows how WI calculates these four indicators for authors. The example concerns the hypothetical profile of an author who has published a total of 3 works. The different versions of these works (editions, translations, reprints, etc.) have generated five publications that are indexed in 159 library catalogs associated with WorldCat. This example, as other authors have pointed out (Zuccala et. al, 2018), indicates that we are not counting books in a physical sense (i.e. items with a unique ISBN) but are measuring intellectual contributions. These contributions are the sum of different types/versions and formats of works, as exemplified in Table 1. In the present paper, we have used these indicators together with each author's citation record taken from Google Scholar profiles and the Web of Science (WoS) Core Collection using the beta Author Search.

**Table 1** Example of the indicators computed for a fictitious author with five books indexed in WorldCat

| Work | Publication | Language | Holdings |
|---|---|---|---|
| Work 1 | 1 ed – country a | Spanish | 10 |
| | 2 ed - country a | Spanish | 2 |
| | 1 ed - country b | English | 12 |
| Work 2 | 1 ed - country a | Spanish | 13 |
| Work 3 | 1 ed - country b | English | 122 |
| **3 works** | **5 publications** | **2 languages** | **159 holdings** |

## 2.3. Retrieving Information from WorldCat Identities

WorldCat Identity information can be retrieved directly from the interface or from the API (Applications Programming Interface). First, we used the WI API[5] to automatically search for each author by name, select the most relevant personal identity result, and retrieve all profile information. However, WI has a disambiguation issue that meant we needed to process this search manually in order to check whether the data was correct and then add records from duplicate identities, recording the URLs of all the author's records. Once we had reviewed and corrected this, we used the API to automatically obtain each author's data on 26 November 2019. The data retrieved were: I) basic information and WorldCat indicators; II) the author's 20 most widely held WorldCat works and related indicators (WI does not provide access to further works) and III) the works' language distributions. Data were then combined by author and all previously-gathered information added. We also identified each author's professional role, status, affiliation, and Google Scholar citation record on 16 December 2019 - the EC3 Scholar Mirrors portal includes this information but it has not been updated. The Python software used to analyze the data is available in a GitHub repository[6].

---

[4]Scholar Mirrors: Methodology [11/02/2020]: http://www.scholar-mirrors.infoec3.es/layout.php?id=methodology
[5] WorldCat Identities API: https://pypi.org/project/worldcatidentities/
[6] https://github.com/Wences91/library_catalog_wi/

2.4. Verification process

A library count indicator is difficult to create because complete lists of holdings are needed in order to identify duplicates. The holdings indicator includes duplicate libraries when the same library holds multiple publications and, for any given author, books they have authored or co-authored, edited or contributed to (e.g., as a chapter author). It also includes works dedicated to the author or about the author, including *festschrifts* in their honor, and theses they have supervised. The results can include non-book sources, such as newspaper interviews with the author or historical letters sent to the author. Profiles sometimes contain mistakes, such as works written by other authors, but the results for the present sample seemed to be largely correct. To check the validity of the profiles, we verified the results for the authors listed in Tables 2 and 3.

For the checks, through the API, we downloaded the works by the author with the highest number of holdings returned by the API and checked them manually. The API returns a maximum of 20 works per profile, but as some authors have more than one record in some cases the works recovered is higher than that limit. A total of 1125 works were checked, 98 (9%) of which were not correctly assigned. At the author level, 20 profiles included at least one wrongly assigned work. However, four authors had more than 10 incorrectly assigned works (Aparna Basu, Paul Wouters, Henry Small and José Luis Ortega). Details of these errors and how they affected the results are in Table 5 and in electronic supplementary material 1. The aforementioned 98 works appear in 1751 library holdings but the authors' total number of holdings is 103 796 (Tables 2 and 3). In other words, at profile level, the final count of holdings has a 1.65% error rate, although this is much higher for some authors.

## 3. Results

### 3.1. Author level analysis

Of the 398 authors drawn from EC3's Scholar Mirrors, 129 researchers were not in WorldCat Identities. These authors were not present for three reasons: a) The authors have not published any book, b) the authors do not have works indexed in any library catalogue (For example, they produce grey literature), c) Library catalogues where books are catalogued are not part of WorldCat. Therefore 269 were in WI and 461 records were recovered, including duplicates. In total, 113 authors had more than one record and 156 had only one. We also excluded four authors we considered not to be closely involved in Scientometrics, giving a final sample of 265 authors distributed in 456 records, which we subsequently classified. In July 2020, 232 of this sample were active, 12 emeritus, 11 had died, and 10 retired. Furthermore, 150 were professors, 70 researchers, 42 librarians, and 3 were professionals in the field. A total of 141 105 mentions have been collected from WorldCat. Electronic supplementary material 2 shows the complete list of authors and indicators used in this study.

**Table 2** Top 25 historical (i.e. emeritus, retired or deceased) authors according to library holdings and WI information

| Author | Main Affiliation | Library Holdings | Works | Holdings / Work | Publications |
| --- | --- | --- | --- | --- | --- |
| Blaise Cronin | Indiana University | 6785 | 144 | 47.12 | 586 |
| Derek J. de Solla Price | Yale University | 6775 | 179 | 37.85 | 484 |
| Jose Maria López Piñero | CSIC | 5551 | 750 | 7.40 | 1836 |
| Eugene Garfield | Institute for Scientific Information | 3399 | 148 | 22.97 | 448 |
| Péter Jacsó | University of Hawaii | 3362 | 25 | 134 | 77 |
| Michael E. D. Koenig | Long Island University | 2871 | 34 | 84.44 | 136 |
| Tibor Braun | Loránd Eötvös University | 2600 | 163 | 15.95 | 452 |
| Alan Pritchard | National Computing Centre (UK) | 2515 | 78 | 32.24 | 199 |
| Vasily V. Nalimov | ---- | 2441 | 83 | 29.41 | 262 |
| Henk F. Moed | Leiden University | 2398 | 49 | 48.94 | 169 |
| Michael J. Moravcsik | University of California | 2225 | 59 | 37.71 | 192 |
| Peter Ingwersen | University of Copenhagen | 1862 | 113 | 16.48 | 270 |
| Howard D. White | Drexel University | 1821 | 18 | 101.17 | 63 |
| Ronald Rousseau | KU Leuven | 1381 | 23 | 60.04 | 119 |
| Yves-François Le Coadic | Cnam – Paris | 1302 | 30 | 43.40 | 91 |
| Loet Leydesdorff | University of Amsterdam | 1234 | 65 | 18.98 | 192 |
| Sven Hemlin | University of Gothenburg | 1124 | 34 | 33.06 | 99 |
| Bertram C. Brookes | University College London | 963 | 47 | 20.49 | 203 |
| Samuel C Bradford | Science Museum London | 721 | 48 | 15.02 | 140 |
| Anthony F.J. van Raan | Leiden University | 439 | 39 | 11.26 | 76 |
| Francis Narin | CHI Research | 422 | 46 | 9.17 | 96 |
| Belver C. Griffith | Drexel University | 352 | 24 | 14.67 | 54 |
| András Schubert | Hungarian Academy of Sciences | 349 | 21 | 16.62 | 62 |
| Aparna Basu* | NISTADS | 297 | 10 | 29.70 | 68 |

*This author's record includes many incorrectly assigned works.

The 265 authors found have an average of 22.4 works and 52.0 publications, in an average of 1.94 languages. The average number of library holdings is 532. In the authors' Google Scholar profiles, the average total number of citations is 3186, of which they received 1651 between 2014 and 2019. There are substantial differences between authors, with the 25 authors with the most works accounting for 49.8% of the total. Although historical authors account for only 12.5% of our sample, they produced 39.5% of the works. To introduce our results, we have divided the authors into two subsets: the historical figures of Scientometrics (Table 2) and currently active figures (Table 3).

The historical researcher with the highest number of library holdings (6785) is Blaise Cronin, from the University of Indiana and a former editor of JASIST. He is followed by Derek J de Solla Price—one of the fathers of Bibliometrics—and José María López-Piñero—who introduced Bibliometrics into Spain—with 6775 and 5562 holdings, respectively. There seem to be no notable authors absent from the table. It includes early contributors like Alan Pritchard (2515)—one of the first to define Bibliometrics in 1969—and the generation of the 1950s and 1960s with Eugene Garfield (3339) or Nalimov (2441)—the father of Science in the Soviet Union and author of *Naukometria*. The list also includes the more recent generation which definitively consolidated the field, with figures like Tibor Braun (2706)—who founded *Scientometrics* in 1978—Henk F. Moed (2398)—one of the first members of the Centre for Science and Technology Studies (CWTS) at Leiden University—and Loet Leydesdorff (1234). It excludes figures that were influential in Scientometrics but were not primarily scientometricians, such as Robert Merton.

**Table 3** Top 25 active authors according to library holdings and WI information

| Author | Main Affiliation | Library Holdings | Works | Holdings / Work | Publications |
|---|---|---|---|---|---|
| Caroline S. Wagner | Ohio State University | 7157 | 32 | 223.66 | 147 |
| Chaomei Chen | Drexel University | 5879 | 42 | 139.98 | 243 |
| Katy Börner | Indiana University Bloomington | 5077 | 46 | 110.37 | 163 |
| Paul Wouters* | Leiden University | 3582 | 60 | 59.70 | 123 |
| Nick Tomaiuolo | Connecticut State University | 3186 | 5 | 637.20 | 36 |
| Ben R Martin | Prof Ben Martin | 3014 | 60 | 50.23 | 192 |
| Peter Van den Besselaar | Vrije Universiteit Amsterdam | 2920 | 52 | 56.15 | 179 |
| Lokman Meho | American University of Beirut | 2420 | 11 | 220.00 | 46 |
| Cassidy R. Sugimoto | Indiana University Bloomington | 2301 | 11 | 209.18 | 88 |
| Andrea Scharnhorst | The DANS KOS Observatory | 2234 | 15 | 148.93 | 67 |
| Ian Rowlands | University of Waterloo | 2090 | 23 | 90.87 | 101 |
| Fiorenzo Franceschini | Politecnico di Torino | 2042 | 28 | 72.93 | 98 |
| Radhamany Sooryamoorthy | University of KwaZulu-Natal | 1876 | 27 | 69.48 | 85 |
| Bart Van Looy | KU Leuven | 1810 | 90 | 20.11 | 175 |
| Koenraad Debackere | KU Leuven | 1637 | 105 | 15.59 | 175 |
| Kim Holmberg | University of Turku | 1617 | 12 | 134.75 | 46 |
| Ying Ding | Indiana University Bloomington | 1374 | 14 | 98.14 | 83 |
| HD Daniel | ETH Zurich | 1351 | 36 | 37.53 | 92 |
| Stefanie Haustein | University of Ottawa | 1317 | 11 | 119.73 | 32 |
| Javier Ruiz-Castillo | Universidad Carlos III | 1186 | 171 | 6.94 | 301 |
| Wolfgang Glänzel | KU Leuven | 1111 | 54 | 20.57 | 114 |
| José Luis Ortega* | CSIC | 1095 | 23 | 47.61 | 50 |
| Svein Kyvik | Nordic Institute for Studies in Innovation | 1063 | 65 | 16.35 | 140 |
| Mustar Philippe | Ecole des Mines de Paris | 1057 | 39 | 27.10 | 112 |
| Mike Thelwall | University of Wolverhampton | 998 | 34 | 29.35 | 114 |

*This author's record includes many incorrectly assigned works.

Table 3 shows those researchers who currently remain active and have not retired. It includes the researcher with the largest number of library holdings in the sample: Caroline S Wagner from Ohio State University with 7157 holdings. There are also two researchers with over five thousand library holdings: Chaomei Chen from Drexel University (5879) and Katy Börner from Indiana University at Bloomington (5077). The list also includes two librarians: Nick Tomaiuolo at Connecticut State University (3186)—who has published just five books mainly related to library management—and Lockman Meho (2421) from Lebanon. At the university level, KU Leuven has the most researchers on the list (Van Looy, Debackere and Glänzel). If we compare Tables 2 and 3, we see that at author level library holdings favor current researchers as almost all of them have at least 1000 library holdings. However, we have only classified 38 as historical, so their averages are higher. Some 59% of active authors have fewer than 50 library holdings and 30% have three or fewer.

To complement this analysis we compared the library holdings data with total citations in Google Scholar profiles, —we used the Spearman correlation rather than the Pearson correlation—and found a weak positive correlation with the number of citations (0.49). A group of authors with higher values for both indicators stands out, but we also find highly cited authors (for example Leydesdorff, Thelwall or Glänzel) with relatively few library holdings for their citations. These authors may focus more on publishing scientific articles and write comparatively few books, hence their lower library holdings. This suggests that these indicators may help to distinguish between highly visible authors with significant book authorship and those who are highly visible overall.

**Fig. 2** Library holding and Google Scholar Citations for main Bibliometrics authors classified by status

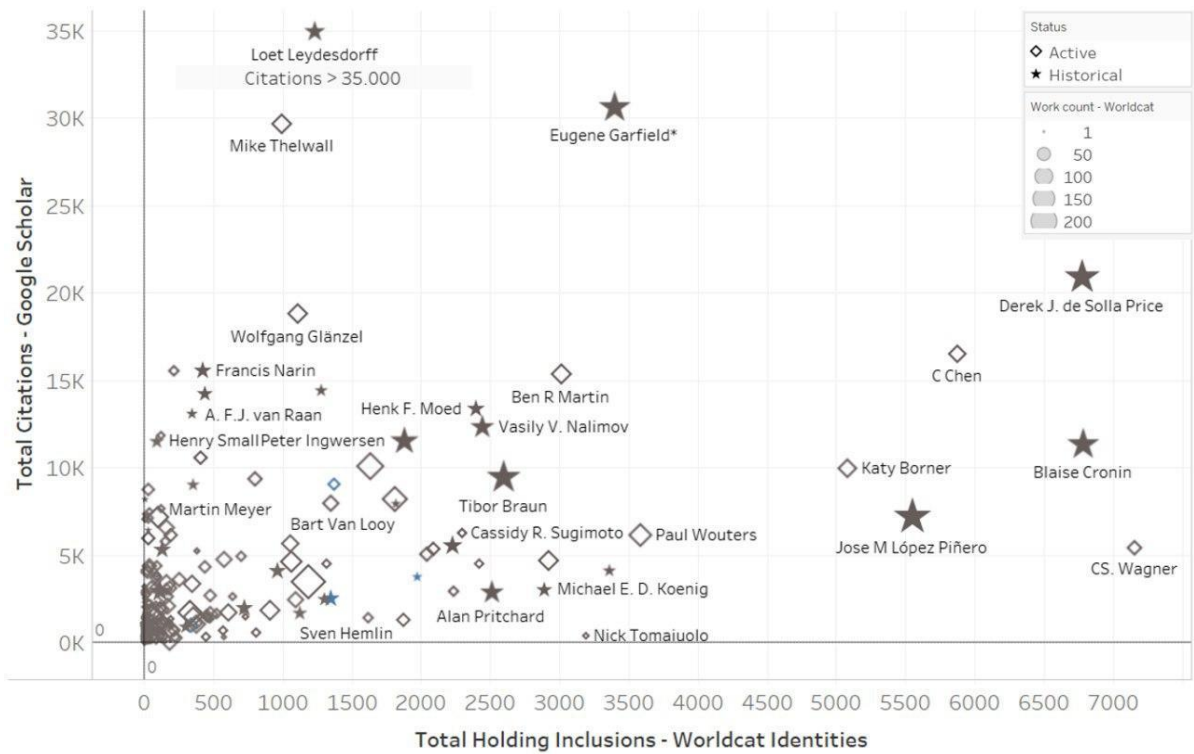

**Fig. 3** Library Holding and Web of Science Core Collection citations for the main Bibliometrics authors classified by status

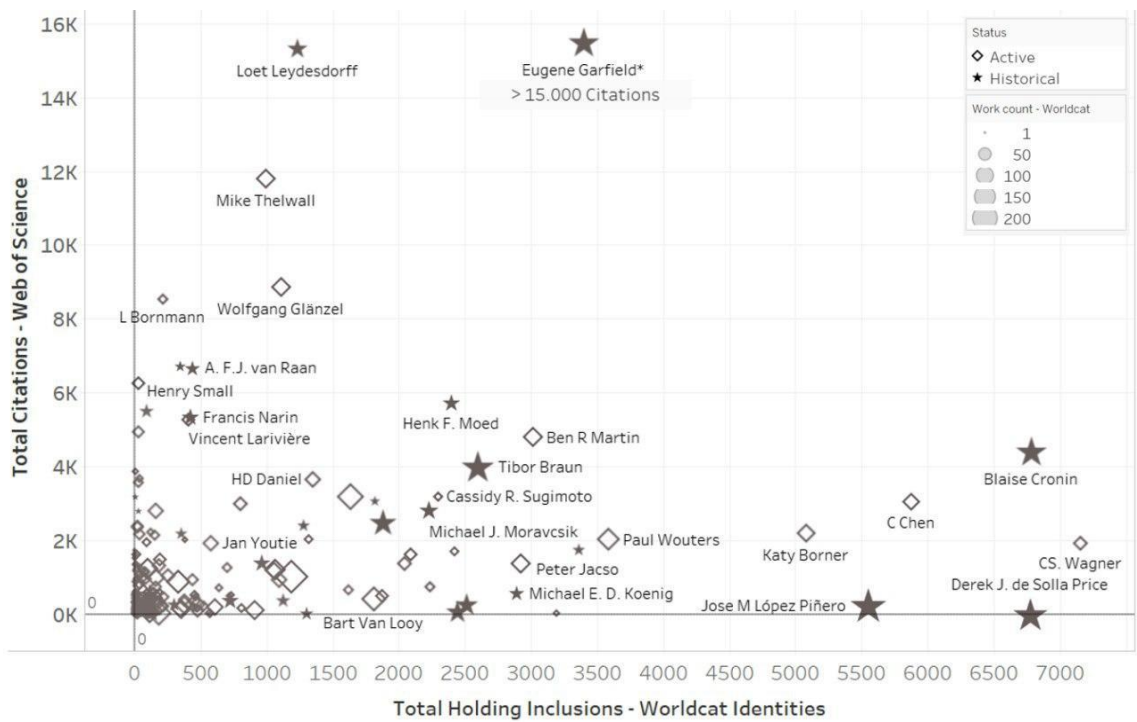

Library holdings were also compared with indicators calculated from the WoS Core Collection. If we compare WoS citations with library holdings, the image we obtain is similar to that in Fig. 2, although the Spearman correlation of 0.22 is weaker (Fig. 3). Google Scholar indexes books, potentially bringing its citation indicator closer to the number for library holdings than that for WoS citations. Fig. 3 shows that some of the most frequently cited researchers today (e.g., Bormann, Lariviere) have no impact on library holdings. Again, we have clear evidence of a researcher profile that is oriented towards journals and ignores books as a channel of communication. The two comparative figures therefore reveal different author profiles and demonstrate the value of using multiple indicators.

### 3.2. Book-level analysis

The authors in our sample have 5925 works and 13 786 publications in WorldCat. However, we have been unable to recover all of them, so our final sample is limited to 3134 works (52.9%) and 9484 publications (68.8%). Our total sample of books was 2668 following a cleaning process that involved eliminating duplicate works caused by co-authorship and books wrongly assigned to the authors being analyzed (Electronic supplementary material 1). Some 223 of these authors have at least one publication in English, while 78 have one in Spanish, and 45 have one in German. Of the works recovered, the most common language is English, accounting for 68.3% of the total, with 13.8% in Spanish and 4.2% in German. Most of the books analyzed (83%) correspond to active authors, and these account for 76% of library holdings. In total, 119 264 books are included in WorldCat and 89 959 are cited on Google Scholar, which means an average of 44 and 33 library holdings per book, respectively. The boxplot distributions of library holdings in Fig. 4 shows that books published by historical authors have higher averages even though the books with higher library holdings figures correspond to active authors. Some of the works are monographs and others are edited volumes. The contribution of an editor is presumably less than the contribution of the monograph author since editors should share credit with chapter authors.

**Fig. 4** Distribution of library holdings in WI at book level

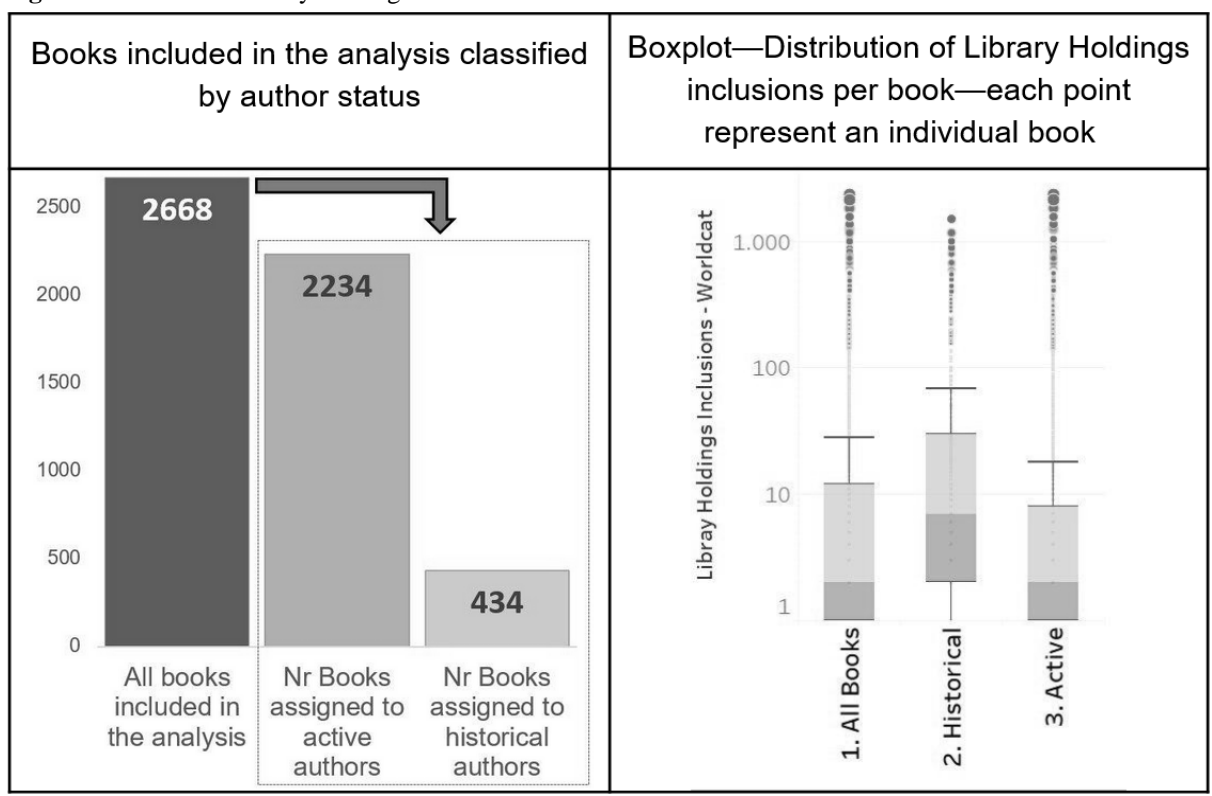

Table 4 ranks books published by authors who specialize in Scientometrics, ordered by the number of library holdings. The first three books have two common characteristics: they have an applied/professional nature and receive few citations. The book that is most visible in catalogs is *Global science & technology information: a new spin on access* by Wagner, with 2378 library holdings and 6 citations. This is followed by *Build your own databases* by Jacsó (2137 libraries, 14 citations) and *The Web library* by Tomaiuolo (1845 libraries and 9 citations). Format seems to have an impact on which books are available. For example, *Global science & technology information: a new spin on access* is listed by WorldCat as available at the University of Wolverhampton (to which one of the present authors is affiliated). However, it is only in electronic format as part of a ProQuest Ebook Central subscription despite being listed in WorldCat as a "Print Book" (which does also exist). This title is part of the UK & Ireland Academic Complete package of 115 000 ebooks provided to UK and Irish universities by ProQuest[7] and, therefore, part of a package selected by ProQuest rather than by a librarian[8]. Similarly, *Compass for intercultural partnerships* and *Theories of informetrics and scholarly communication: a* Festschrift *in honor of Blaise Cronin* are part of the ProQuest Academic Complete ebook offering for the USA, presumably accounting for their high library holdings.

Table 4 Ranking of books with a higher number of Library Holdings

| Book. Bibliographic reference | Library holdings | Google Scholar Citations |
|---|---|---|
| Caroline Wagner, Allison Yezril. **Global science & technology information: a new spin on access**. RAND, 1999 | 2378 | 6 |
| Péter Jacsó; F Wilfrid Lancaster. **Build your own database**. American Library Association, 1999. | 2137 | 14 |
| Nicholas G Tomaiuolo; Barbara Quint. **The Web library: building a world class personal library with free Web resources. Information Today**, 2004. | 1845 | 9 |
| Katy Börner. **Atlas of knowledge: anyone can map**. The MIT Press, 2015 | 1565 | 97 |
| Steven W Popper; Eric V Larson; Caroline S Wagner. **New forces at work: industry views critical technologies**, RAND, 1998 | 1557 | 88 |
| Derek de Solla Price. **Science since Babylon.** Yale Univ. Press 1962 | 1515 | 1381 |
| R Elsen; Ignace Pollet; Patrick Develtere; Koenraad Debackere. **Compass for intercultural partnerships**. Leuven University Press, 2017 | 1371 | 2 |
| Paul Wouters; Anne Beaulieu; Andrea Scharnhorst; Sally Wyatt. **Virtual knowledge: experimenting in the humanities and the social sciences**. MIT Press, 2013 | 1342 | 47 |
| Nicholas G Tomaiuolo. **UContent: the information professional's guide to user-generated conten**t. **Information Today**, 2012 | 1336 | 6 |
| Stefanie Haustein **Multidimensional journal evaluation: analyzing scientific periodicals beyond the impact factor.** De Gruyter/Saur, 2012. | 1305 | 52 |
| Katy Börner; David E Polley. **Visual insights: a practical guide to making sense of data**. MIT Press, 2014. | 1289 | 107 |
| Caroline S. Wagner. **The new invisible college: science for development**. Brookings Institution Press, 2008. | 1269 | 547 |
| Chris Steyaert; Bart van Looy. **Relational practices, participative organizing**. Emerald, 2010. | 1255 | 28 |
| Howard D. White. **Brief tests of collection strength: a methodology for all types of libraries**. Greenwood Press, 1995. | 1179 | 60 |
| Henk F Moed. **Citation analysis in research evaluation**. Springer, 2005 | 1015 | 1171 |
| Vladimir Geroimenko; Chaomei Chen. **Visualizing the semantic Web: XML-based Internet and information visualization.** Springer, 2003 | 948 | 294 |
| Cassidy R Sugimoto; Blaise Cronin. **Theories of informetrics and scholarly communication: a Festschrift in honor of Blaise Cronin**. De Gruyter, 2016. | 921 | 12 |
| Derek de Solla Price. **Frontiers of science: on the brink of tomorrow**. The Society, 1982. | 894 | -- |
| Péter Jacsó. **Content evaluation of textual CD-ROM and Web databases**. Libraries Unlimited, 2001. | 872 | 16 |

In terms of content, these books are not necessarily oriented towards bibliometric issues and many focus instead on technological issues (web design, databases, etc.). Books of a combined professional and educational nature dominate the table. This list also includes at three books with both a high number of citations and a high number of library holdings. Unlike those featured earlier, these books were written by the historical authors of the field and their contents are not specialized in Scientometrics in two cases. The first is the classic foundation book *Science since Babylon* which describes the exponential growth of scientific literature. It is fifth in the library holdings ranking (1515) and is the most cited book in our collection (1381). It is followed by a title with a similar profile: Wagner's *New invisible college: science for development* (1269 library and 547 citations) and *Citation analysis in research evaluation* by Henk F. Moed (1015 libraries and 1171 citations). Table 4 therefore reveals two distinct book profiles: those with a Scientometric focus and those focusing more on the sociology of science.

---
[7] https://about.proquest.com/blog/pqblog/2015/Brand-New-UKI-Edition-of-Academic-Complete.html
[8] See also: https://about.proquest.com/products-services/Academic-Complete.html

## 4. Discussion & Conclusions

This paper has investigated WorldCat Identities for comparative analyses of authors through their library holdings. We investigated the viability of applying this source of information to a specific scientific field. The investigation shows that this type of analysis is possible but that WI has a series of methodological limitations (Table 5). Most of the problems described relate to the incorrect disambiguation of author names and the incorrect assigning of works. This means we cannot directly use the indicators given to an author without first verifying and validating the data collected. When using the API directly, we successfully located 221 authors (82.2% of the total collected in WI). In addition, 125 of these authors had no duplicate records (46.6% percent of the total and 80.1% of the total of non-duplicated records). Furthermore, we recovered 96 authors but found duplicate records (85.0% of the authors had duplicates; 31.5% of the records were duplicated).

**Table 5** List of main methodological limitations of WI

| Limitation | Description |
| --- | --- |
| Does not disambiguate well in Spanish | Does not disambiguate homonyms, thus sometimes generating multiple entries. |
| Does not aggregate authors from different sources | Many authority records are generated from other authority sources, such as VIAF or LCCN, but the authors who are present in several of them are not unified in WI. |
| Separation of personal and corporate identity | Some authors have separate records for personal identity and corporate identity. |
| Does not only include books | Includes materials such as theses, articles or conference papers and proceedings. |
| Incorrect assignment of records | Authors may have other authors' work associated with their records. |
| Conflict between works written by the author and works about the author | WI differentiates between the author's own work and works about the author—such as biographies—although these are sometimes confused. |
| Includes catalog entries from large scale ebook subscriptions | Books can have high values if they are included in a version of ProQuest Academic Complete or any other ebook service that integrates with library catalogs. |

Furthermore, database use can involve geographic or linguistic bias, both of which are very common in citation indexes. Despite WorldCat's obvious advantages, few studies have critically analyzed its coverage even though it has a clear English language bias (Wakeling, Clough, Connaway, Sen & Tomás, 2017). Table 6 shows that 44.8% of platform users and 43% of academic libraries are from the US—much higher figures than those for any European country. For example, only 1.5% of users and 0.7% of libraries are in Spain—of 76 Spanish university libraries, only 42 are in WorldCat. Thus, Spanish researchers may need to use complementary sources when analyzing the diffusion of books in Spain. Thus, when conducting an LCA with WorldCat, researchers should consult the OCLC members' directory to verify the catalogs' territorial distribution. In addition, Table 6 may allow us to take an objective approach to the significance and the terminological debate. In this paper we use the terms 'impact' as well as 'book diffusion' for library catalog measures but the term 'cultural presence' or 'cultural visibility' (Zuccala, 2018) may be more precise. The results have shown that the vast platforms and users in WorldCat are from the United States and this could be a signal of this cultural 'availability', 'presence' or 'visibility'.

**Table 6** User location and number of academic libraries in WorldCat (OCLC members only)

| Countries | User Location[note1] | Number of academic libraries[note2] | Number of public libraries[note2] | Number of other types of library[note2] | Total number of libraries[note2] |
|---|---|---|---|---|---|
| United States | 44.8 | 2198 (40.69 %) | 3394 (80.79 %) | 3264 (76.96 %) | 8856 (63.97 %) |
| China | 5.3 | 13 (0.24 %) | 8 (0.19 %) | 4 (0.09 %) | 25 (0.18 %) |
| Canada | 5.2 | 117 (2.17 %) | 73 (1.74 %) | 113 (2.66 %) | 303 (2.19 %) |
| United Kingdom | 3.7 | 72 (1.33 %) | 96 (2.29 %) | 82 (1.93 %) | 250 (1.81 %) |
| Germany | 3.2 | 274 (5.07 %) | 16 (0.38 %) | 75 (1.77 %) | 365 (2.64 %) |
| France | 2.3 | 1076 (19.92 %) | 8 (0.19 %) | 29 (0.68 %) | 1113 (8.04 %) |
| India | 1.8 | 23 (0.43 %) | 0 (0 %) | 11 (0.26 %) | 34 (0.25 %) |
| Italy | 1.7 | 87 (1.61 %) | 112 (2.67 %) | 12 (0.28 %) | 211 (1.52 %) |
| Indonesia | 1.7 | 32 (0.59 %) | 0 (0 %) | 42 (0.99 %) | 74 (0.53 %) |
| Spain | 1.5 | 15 (0.28 %) | 4 (0.1 %) | 14 (0.33 %) | 33 (0.24 %) |
| Netherlands | 1.5 | 38 (0.7 %) | 95 (2.26 %) | 24 (0.57 %) | 157 (1.13 %) |
| Mexico | 1.3 | 24 (0.44 %) | 0 (0 %) | 5 (0.12 %) | 29 (0.21 %) |
| Australia | 1.3 | 156 (2.89 %) | 232 (5.52 %) | 376 (8.87 %) | 764 (5.52 %) |
| Brazil | 1.3 | 17 (0.31 %) | 0 (0 %) | 3 (0.07 %) | 20 (0.14 %) |
| Poland | 1.2 | 17 (0.31 %) | 4 (0.1 %) | 3 (0.07 %) | 24 (0.17 %) |
| Japan | 0.9 | 64 (1.18 %) | 1 (0.02 %) | 12 (0.28 %) | 77 (0.56 %) |
| Malaysia | 0.9 | 23 (0.43 %) | 0 (0 %) | 3 (0.07 %) | 26 (0.19 %) |
| Korea, Republic of | 0.7 | 6 (0.11 %) | 0 (0 %) | 1 (0.02 %) | 7 (0.05 %) |
| Russian Federation | 0.7 | 13 (0.24 %) | 0 (0 %) | 9 (0.21 %) | 22 (0.16 %) |
| Singapore | 0.7 | 12 (0.22 %) | 2 (0.05 %) | 11 (0.26 %) | 25 (0.18 %) |

[note1] Geographical location of users, results from log (Wakeling, Clough, Connaway, Sen & Tomás, 2017)
[note2] Data from the Directory of OCLC members

As with other scientific databases, WorldCat has common information retrieval and coverage problems that must always be borne in mind when conducting a study. Nonetheless, despite these limitations, WI is a relevant source for studies of authors. In this article, we have analyzed the field of Scientometrics and our results include the most frequently cited researchers in the field, both historical and current. The results confirm that authors have different publication profiles so that focusing on journal articles alone may disadvantage book-oriented scholars. For example, highly cited authors like Price and Wagner have high library holdings numbers whereas others, like Leydesdorff, have library holdings numbers that do not correspond to them and, finally, other authors, like the librarian Nick Tomaiuolo, have few citations and high library holdings numbers.

Library holdings are most relevant to authors or editors of handbooks, monographs, or textbooks. This classification reflects a different sphere of activity and academic contributions related to the generation of teaching/educational contents (e.g. textbooks) or the author's engagement in their field (e.g. editing conference proceedings). Clearly some authors contribute to a specific field as well as undertaking other activities or publishing materials other than articles. In this context, both Chaomei Chen and Blaise Cronin serve as examples. For example in Chen's profile two of his most relevant contributions are textbooks oriented towards undergraduate and graduate students (*Visualizing information using SVG and X3D* and *Information visualization: beyond the horizon*). In the case of Cronin's profile, some works have a professional and/or educational approach (*The marketing of library and information services*) and even some with a clear humorous entertainment component (*Pulp friction*). We have also detected profiles that are 100% professional, especially those of librarians who have remained outside of scientific circles. Our classification captures the value of an academic activity beyond journal citations.

We have also analyzed the most relevant books and this has allowed us to discuss the previous results and distinguish between two contrasting phenomena. Firstly, we have a set of books with great scientific impact and diffusion in libraries. These include foundational texts in the field like *Science since Babylon* (Price, 1962) and contemporary manuals such as *Citation analysis in research evaluation* (Moed, 2005). These books enjoy widespread scientific recognition and, moreover, are reference manuals or handbooks—a value encapsulated in the library counts indicator. Secondly, we have a set of books that are present in many libraries but which have few citations. They may have a practical, professional profile, are not oriented towards a scientific readership and, thanks to the library counts, can now be analyzed from a different standpoint. One example of this profile is *Build your own database* (Jacsó & Lancaster, 1999). The list of works also shows that many authors in the field of Scientometrics publish non-specialized works that are more of a professional or educational nature and which go unnoticed in the more traditional bibliometric analyses.

The list of books should be interpreted in the context of its limitations. With our methodology we have identified a large number of books with the highest "Library Catalog Counts". This methodology does not identify all important books, however. For example, if an important book's author is not in EC3's Scholar Mirror, that book would not be in our study. Therefore, the list is likely to be incomplete. Another issue that could question the value and significance of this list is the fact that books found in a catalog do not always respond to a librarian's choice since some are donations (Biagetti, 2018). Book holdings are primarily the result of decisions made by collection librarians, with the exception of big deal packages, which are collated by publishers. Thus, the library holding results are indirect indicators of impact or diffusion in the sense that they rely on the channels of information available to collection librarians to make judgements about the types of books that they believe to be relevant to their audience. However, perhaps the factor that most distorts the value of libcitations—as a consequence of the selection process—is the purchase of e-book collections, since these integrate books into library collections *en masse* (Lewis & Kennedy, 2019). In our case the holdings counts for some books are substantially boosted by their presence in ProQuest (or other) mass electronic offerings. This substantially undermines the value of library holdings as indicators of academic interest in books because they do not always reflect the decisions of individual librarians and academics when purchasing books, even though they do reflect the availability of books in libraries.

Finally, whilst WI may be useful for indicators based on library holdings, the limitations above should be taken into account when using them or the results may be highly misleading. Like most current metric profiles, the indicators have to be reviewed and corrected. This problem has been reported, for example, for Google Scholar, ResearchGate and Mendeley profiles (Martín-Martín, Orduña-Malea & Delgado López-Cózar, 2016) although the advantage of WI is that, unlike other sources, it cannot be manipulated (Thelwall & Kousha, 2017). In addition, library holdings are influenced by the presence of a book in a package deal, such as that of ProQuest, and there does not seem to be a practical way to detect this (there does not seem to be a public ProQuest list of Academic Complete that could be checked, for example). Nevertheless, WI may be useful tool because at author level it offers quickly-obtained indicators that allow us to present an alternative vision of impact even though it must be used with checks to safeguard against inflation due to books in package deals.